\documentclass[%
reprint,
amsmath,amssymb, aps,
]{revtex4-1}

\usepackage{graphicx}% Include figure files
\usepackage{dcolumn}% Align table columns on decimal point
\usepackage{bm}% bold math
\usepackage{braket}
\usepackage{amsmath}
\begin{document}

    %\preprint{APS/123-QED}

    \title{New phase space calculations for $\beta$-decay half-lives}% Force line breaks with \\
    \author{Sabin Stoica$^{1,2}$, Mihail Mirea$^{1,2}$, Ovidiu Ni\c{t}escu$^{2,3}$}
    \affiliation{ $^1$Horia Hulubei Foundation, P.O. MG6, 077125-Magurele,
        Romania, $^2$Horia Hulubei National Institute of Physics and Nuclear
        Engineering, P.O. Box MG6, 077125-Magurele, Romania,\\
        $^3$University of Bucharest, Faculty of Physics, P.O. Box MG11, 077125-Magurele, Romania}
    \email{stoica@theory.nipne.ro} \altaffiliation[Also at ]{Horia
        Hulubei Foundation, P.O. MG12, 077125-Magurele, Romania.}
    \author{Jameel-Un Nabi}
    \affiliation{GIK Institute of Engineering Sciences and Technology,
        Topi 23640, Khyber Pakhtunkhwa, Pakistan.}
    \author{Mavra Ishfaq}
    \affiliation{GIK Institute of Engineering Sciences and Technology, Topi 23640,\\
        Khyber Pakhtunkhwa, Pakistan.}
    \date{\today}

    \begin{abstract}

We revisit the computation of the phase space factors (PSF) involved
in the positron decay and electron capture (EC) processes for a
large number of nuclei of experimental interest. To obtain the
electron/positron wave functions needed in computation, we develop a
code for solving accurately the Dirac equation with a nuclear
potential derived from a realistic proton density distribution in
the nucleus. The finite nuclear size (FNS) and screening effects are
included through recipes which differ from those used in previous
calculations. Comparing our results with former calculations
employing approximate methods but computed with the same Q-values,
we find a close agreement for positron decays, while for the EC
process there are relevant differences. For the EC process we also
find that the screening effect has a notable influence on the
computed PSF values specially for light nuclei. Further, we
re-computed the same PSF values but using the most recent Q-values
reported in literature. In several cases these new Q-values differ
significantly from the older ones, which results in large
differences in the PSF values as compared with previous results.
These new PSF values proposed here, can contribute to a more
reliable calculation of the beta decay rates, which are key
quantities in the study of nuclei far from the stability line, as
well as to better understanding of the stellar evolution.

        \begin{description}
            \item[PACS numbers]23.40.Bw; 23.40.-s;  26.30.Jk

        \end{description}
    \end{abstract}

    \pacs{Valid PACS appear here}% PACS, the Physics and Astronomy
    % Classification Scheme.
    %23.40.Bw    Weak-interaction and lepton (including neutrino) aspects
    %(see also 14.60.Pq Neutrino mass and mixing)
    % 23.40.-s    ß decay; double ß decay; electron and muon capture
    % 26.30.Jk    Weak interaction and neutrino induced processes, galactic radioactivity

    \maketitle

    %\tableofcontents

    \section{\label{sec:level1}INTRODUCTION}

The phase space factors for beta decay and electron capture were
calculated since long time \cite{Beh69, Mar70,Gov71} and were
considered to be evaluated with sufficient accuracy. However, in
those works the distortion of the electron wave functions (w.f.) by
the Coulomb field of the nucleus was taken into account through
Fermi functions which were expressed in terms of approximate radial
solutions of the Dirac equation at the nuclear surface. Also, other
corrections were introduced in the calculations in approximate ways.
Thus, the screening effect on the $\beta$ spectrum was included by
various recipes, for example by replacing the V(Z) potential with a
momentum dependent screening (for low energy positrons)
\cite{Gov71}, by modifying the electron radial w.f.
\cite{Ro36}-\cite{G54}, etc. Also, the finite size of the nucleus (FNS) was
taken into account by adding to the Fermi functions obtained in the
"point-nucleus" approximation, corrections that depend on the
$\beta$ particle energy and nuclear charge Z \cite{RH51,RH57}. Also,
for the nuclear radius, older formula has been used
\cite{Gov71,El61}. For the EC process the electron bound-state
radial w.f. were also obtained as approximate solution of the Dirac
equation evaluated at the nuclear surface. They were improved by
including exchange and overlap corrections, which were obtained
within a relativistic HF approach.

In this work we revisit the computation of the PSF involved in the
positron decay and electron capture (EC) processes for light and
heavy nuclei of experimental interest. The Dirac equation is solved
numerically with a Coulomb potential derived from a realistic proton
distribution in the nucleus which includes the FNS correction. The
numerical procedure follows the power series method described in
Ref. \cite{Buh65} and is similar to that described in Refs.
\cite{sal1,sal2}. The screening effect was introduced by using a
screened Coulomb potential, obtained by multiplying the Coulomb
potential by a function $\phi(x)$, solution of the Thomas-Fermi equation
obtained by the Majorana method \cite{Esp02}. The accuracy imposed
in our numerical algorithms used to solve the Dirac equation always
exceeds the convergence criteria given in those references. Also, a
more efficient procedure to identify  the electron bound states
without ambiguity was developed.

In order to make a comparison between the actual PSF values found in
literature and ours, the same PSF are also computed with the
approach described in Refs. \cite{Mar70,Gov71} and using the same
Q-values. For positron decays our results are in close agreement
with the other previous results, while for the EC process we found
significant differences. For these processes we also find that the
screening effect has a notable influence on the computed PSF values
for light nuclei. Further, we re-computed the same PSF values using
up-dated Q values, reported recently in literature \cite{Aud12},
which for several light nuclei differ significantly from the older
ones. As an example we cite the maximum $\beta$-particle energy
(referred to as $W_{0}$ throughout this paper) stated in Table~2 of
Ref. \cite{Wil74}. These $W_{0}$ values differ considerably from
those given in Ref. \cite{Aud12}. One reason for this big difference
could be that Wilkinson $\&$ Macefield, in order to compare their
calculation with those performed earlier by Towner $\&$ Hardy
\cite{TH73}, restricted their phase space to only pure Fermi
transitions. In other words the Gamow-Teller window was not accessed
in phase space calculation of \cite{Wil74}. Thus, in this paper we
propose new PSF values computed with a more accurate method and
using up-dated Q-values, for a large number of nuclei of
experimental interest. Our calculations can be useful for more
reliable computation of the beta decay rates of nuclei far from the
stability line, as well as for better understanding of the stellar
evolution.

Our work is further motivated by similar calculations done for the
double-beta decay (DBD) process. The PSF for DBD were also
considered for a long time to be computed with enough accuracy and
were used as such for predicting DBD lifetimes. However, recently,
they were recalculated with improved methods, especially for
positron and EC decay modes \cite{Kot12,Mir15} and several
differences were found as compared to previous calculations where
approximate electron/positron w.f. were used.

The paper is organized as follows. In Section II we present briefly
the two approaches used to compute our PSF values. Our results are
reported in Section III. Here we compare them with experimental data
and previous results and discuss the differences. Finally, we
summarize the main points and present our conclusions in Section IV.

    \section{\label{sec:level2}FORMALISM}

Following essentially the formalism from Ref. \cite{Gov71}, we give
here the necessary equations which we use to calculate the PSF.

    \subsection{Phase space factors for $\beta^+$ transitions}

The probability per unit time that a nucleus with atomic mass A and
charge Z decays for an allowed $\beta$-branch is given by:

    \begin{eqnarray}\label{pro}
        \lambda_{0} = g^{2}/ 2\pi^{3}\int_1^{W_0} pW(W_0-W)^2S_{0}(Z, W)dW,
    \end{eqnarray}
where $g$ is the weak interaction coupling constant, $p$ is the
momentum of $\beta$-particle, $W$ = $\sqrt {{p^{2} + 1}}$ is the
total energy of $\beta$-particle and $W_0$ is the maximum
$\beta$-particle energy. $W_0$ = $Q - 1$, in $\beta^{+}$decay ($Q$
is the mass difference between initial and final states of neutral
atoms). Eq.~(\ref{pro}) is written in natural units ( $\hbar = m = c
= 1$ ) so that the unit of momentum is $mc$, the unit of energy is
$mc^{2}$, and the unit of time is $\hbar$ /$mc^{2}$. The shape
factors $S_{0}(Z, W)$ for allowed transitions which appear in
Eq.~(\ref{pro}) are defined as:

    \begin{eqnarray}\label{sf}
        S_{0}(Z, W) = \lambda_1 (Z,W) |M_{0,1}|^2,
    \end{eqnarray}
where $M_{0,1}$ are the nuclear matrix elements and the Fermi
functions $\lambda_1 (Z,W)$. Thus, for calculating the $\beta^+$
decay rates one needs to calculate the nuclear matrix elements and
the PSF,
    that can be defined as:
    \begin{equation}\label{ps1}
        F_{BP}=\int_1^{W_0} pW(W_0-W)^2\lambda_1(W)dW.
    \end{equation}

For the allowed $\beta$ decays the Fermi functions are expressed as:
    \begin{eqnarray}\label{bc}
        \lambda_{1}(Z,W) = {g^2_{-1} + f^2_{1}\over 2p^2},
    \end{eqnarray}
where $g_{-1}(Z,W)$ and $f_1(Z,W)$ are the large and the small
radial components of the positron radial wave functions evaluated at
the nuclear radius $R$ which can be obtained by solving the Dirac
equation:

    \begin{eqnarray}
        \label{dirac} \left(\frac{d}{dr} + \frac{\kappa}{r}\right)g_{\kappa}(W,r) =
        (W+V+1)f_{\kappa}(W,r)\\
        \left(\frac{d}{dr} + \frac{\kappa}{r}\right)f_{\kappa}(W,r)= -(W+V-1)g_{\kappa}(W,r)
        \nonumber
    \end{eqnarray}
where $V$ is the central potential for the positron and
$\kappa=(l-j)(2j+1)$ is the relativistic quantum number. We note
that Eq.~(\ref{dirac}) is also written in natural units.

An important step in the PSF calculation for $\beta^+$ decay is the
method of obtaining the positron continuum radial functions. For
this we develop a new method (code) of solving the Dirac equation,
which is adapted from the method used previously for the computation
of PSF for DBD process \cite{Sab13,Mir15}.

We solved Eq.~(\ref{dirac}) in a nuclear potential $V(r)$ derived
from a realistic proton density distribution in the nucleus. This is
done by solving the Schrodinger equation with a Woods-Saxon
potential. In this case:

        \begin{equation}
            V(Z,r)=\alpha\hbar c\int{\rho_e(\vec{r'})\over \mid
                \vec{r}-\vec{r'}\mid} d\vec{r'} \label{vpot},
        \end{equation}

where the charge density is

        \begin{equation}
            \rho_e({\vec{r}})=\sum_i (2j_i+1)v_i^2
            \mid\Psi_i(\vec{r})\mid^2,
        \end{equation}
$\Psi_i$ is the proton (Woods-Saxon) w.f. of the spherical single
particle state $i$ and $v_i$ is its occupation amplitude. The factor
$(2j_i+1)$ reflects the spin degeneracy.

The screening effect is taken into account by multiplying the
expression of $V(r)$ with a function $\phi(r)$, which is the
solution of the Thomas Fermi equation: $d^2\phi/dx^2 =
\phi^{3/2}/\sqrt x$, with $x=r/b$, $b\approx 0.8853a_0Z^{-1/3}$ and
$a_0$ = Bohr radius. It is calculated within the Majorana method
\cite{Esp02}. The boundary conditions are $\phi(0)=1$ and
$\phi(\infty)=0$. As mentioned above the screening effect is taken
into account by a method developed in Ref. \cite{Esp02}. The
possible ways in which the screening function modifies the Coulomb
potential depends on the specific mechanism and its boundary
conditions.

For the case of the $\beta^+$-decay process, the potential used to
obtain the electron w.f. is

        \begin{equation}
            rV_{\beta^+}(Z,r)=(rV(Z,r)+1)\times\phi(r)-1
        \end{equation}
to take into account the fact that $\beta$ decay releases a final
negative ion with charge -1. $V(Z,r)$ is positive. In our approach,
we considered the solution of the Thomas-Fermi equation as a
universal function, giving an effective screening. Here, the product
$\alpha\hbar c$ = 1, for atomic units. The asymptotic potential
between an positron and an ionized atom is $rV_{\beta^+}=-1$. In
this case, the charge number $Z=Z_0-1$ corresponds to the daughter
nucleus, $Z_0$ being the charge number of the parent nucleus.
Asymptotically $\phi(r)$ tends to zero.

In this case the radial solutions of the Dirac equations should be
normalized  in order to have the following asymptotic behavior

    \begin{equation}\label{asi}
        \begin{split}
            \left(\begin{array}{l}g_{k}(\epsilon,r)\\f_{k}(\epsilon,r)\end{array}\right)\sim
            {\hbar e^{-i\delta_k}\over pr}\\
            \left(\begin{array}{l}\sqrt{{\epsilon+m_ec^2\over 2\epsilon}}
                \sin(kr-l{\pi\over 2}-\eta\ln (2kr)+\delta_k)\\
                \sqrt{{\epsilon-m_ec^2\over 2\epsilon}} \cos(kr-l{\pi\over
                    2}-\eta\ln (2kr)+\delta_k)\end{array}\right),
        \end{split}
    \end{equation}
where $c$ is the speed of the light,  $m_e/\epsilon$ are the
electron mass/energy, $k={p/\hbar}$ is the electron wave number,
$\eta=Ze^2/\hbar v$ (with $Z=\pm Z$ for $\beta^{\mp}$ decays), is
the Sommerfeld parameter, $\delta_\kappa$ is the phase shift and $V$
is the Coulomb interaction energy between the electron and the
daughter nucleus.

On the other side we also calculated the PSF for positron decays
with the method described in \cite{Gov71}. The $g_k$ and $f_k$
functions were calculated by solving the Dirac equation for a
point-nucleus unscreened Coulomb potential, for which the equation
has analytical solutions. The finite nuclear size and screening
effects were introduced as corrections, after the recipe described
in \cite{Gov71}. The finite size correction was introduced by means
of an empirical deviation that depended on the atomic mass $Z$ and
the energy $W$ \cite{RH51,RH57}. The screening correction was given
by the following replacement \cite{Ro36,G54}:
\begin{equation}
            g_{-1}^{2}(Z,W) \rightarrow
            \frac{pW^{\prime}}{p^{\prime}W}g_{-1}^{2}(Z,W^{\prime})
            \nonumber
\end{equation}
\begin{equation}
            f_{-1}^{2}(Z,W) \rightarrow
            \frac{pW^{\prime}}{p^{\prime}W}f_{1}^{2}(Z,W^{\prime}),
\end{equation}
where $W^{\prime} = W + V_{0}$, $p^{\prime} = \sqrt{(W^{\prime})^2 -
1}$ and $V_{0}$ was taken as a $p$-dependent screening potential.
For further details of this formalism we refer to \cite{Gov71}. No
electromagnetic corrections were undertaken in this calculation of
PSF.

\subsection{Phase space factors for electron capture (EC)}
Electron capture is always an alternate decay mode for radioactive
isotopes that do not have sufficient energy to decay by positron
emission. This is a process which competes with positron decay. In
order for electron capture leading to a vacancy in, say, the
K-shell, the atomic mass difference between initial and final
states, $Q$, must be greater than the binding energy of a K-shell
electron in the daughter atom, $\epsilon_K$. The energy carried off
by the neutrino is then given by

    \begin{eqnarray}
        q_K = Q - \epsilon_K
    \end{eqnarray}

If the energy requirement $Q$ $>$ $\epsilon_K$ is satisfied,
electron capture from the K-shell is more probable than that from
any other shell because of the greater density at the nucleus of the
K-shell electrons. The total K-shell capture rate can be expressed
as

    \begin{eqnarray}\label{cr}
        \lambda_{EC,K}^{0} = \lambda^0_K B_K,
    \end{eqnarray}
    where
    \begin{eqnarray}\label{acr}
        \lambda^0_K = {g^2|M_{0,1}|^2 \over 4\pi^2} q^2_K g^2_K,
    \end{eqnarray}
where $g^2$ is a constant (with dimensions of $t^{-1}$), the M's are
specific combinations of nuclear matrix elements, $g_K$ is the large
component of the bound-state radial w.f. of the captured K-shell
electron (evaluated at the nuclear surface $R_{A}$), $q_K$ is the
neutrino energy in units of $mc^2$ and $B_K$ is the "exchange"
correction factor for the K-shell. In analogy with Eq.~(\ref{cr}),
the L-shell total capture rate will be

    \begin{eqnarray}
        \lambda_{EC,L_{i}}^{0} = \lambda^{0}_{L_{i}} B_{L_{i}},
    \end{eqnarray}
where $L_i$ denotes a particular L-subshell. The contribution of
$L_1$ pertaining to the 2s$_{1/2}$ orbital is the most important, so
we keep in our calculations only the contribution of this subshell
to the calculated our PSF. The expressions for
$\lambda^0_{L_{1}}$ can be obtained from Eq.~(\ref{acr}) by the
replacement of $q_K$, $g_{K}$ by $q_{L_{1}}$, $g_{L_{1}}$. Electron
capture from the M-, N- and higher shells may be defined in a
similar fashion, but they have negligible contributions in
comparison with the K- and L- ones.

Hence,  for an allowed transition, the PSF expression of electron
capture within the approximation stated above, can be written as

    \begin{eqnarray}\label{eckl}
        F_{EC}^{K,L_1} = {\pi\over 2} \left(q^2_K g^2_K B_K + q^2_{L_1} g^2_{L_1}
        B_{L_1}\right).
    \end{eqnarray}

    For the $q_{K/L_1}$ quantities we used the expression

    \begin{eqnarray}
        q_{K/L_1}=W_{EC}-\epsilon_{K/L_1},
    \end{eqnarray}
were, $W_{EC}$ is the Q value of the $\beta^+$ decay in $m_ec^2$
units, $\epsilon_i$ are the binding energies of the 1s$_{1/2}$ and
2s$_{1/2}$ electron orbitals of the parent nucleus, $g_i$ their
radial densities on the nuclear surface. $B_i\approx 1$ represent
the values of the exchange correction. These are due to an imperfect
overlap of the initial and final atomic states caused by the one
unit charge difference \cite{Ba63}. In our method we consider these
exchange corrections to be unity, for the nuclei considered, the
estimated error in doing that being under 1\%. The relation
$W_0=W_{EC}-1$ holds.

The $g_{K/L_1}$ are the electron bound states, solutions of the
Dirac equation (\ref{dirac}), and correspond to the eigenvalues
$\epsilon_{n}$ ($n$ is the radial quantum number). The quantum
number $\kappa$ is related to the total angular momentum
$j_\kappa=\mid\kappa\mid-1/2$. These w.f. are normalized
such that

    \begin{equation}
        \int_0^\infty [g^2_{n,\kappa}(r)+f^2_{n,\kappa}(r)]dr=1.
    \end{equation}
For simplicity, we consider solutions of the Dirac equations
$g_{n,\kappa}$ and $f_{n,\kappa}$ that are divided by the radial
distance $r$. An asymptotic solution is obtained by means of the WKB
approximation and by considering that the potential $V$ is
negligible small:

    \begin{equation}
        {f_{n,\kappa}\over g_{n,\kappa}}={c\hbar\over
            \epsilon+m_ec^2}\left({g'_{n,\kappa}\over g_{n,\kappa}}+{\kappa\over
            r}\right),
    \end{equation}
where

    \begin{equation}
        {g'_{n,\kappa}\over g_{n,\kappa}}=-{1\over
        2}\mu'\mu^{-1}-\mu,
    \end{equation}
with

    \begin{equation}
        \mu=\left[{\epsilon+m_ec^2\over \hbar^2
            c^2}(V-\epsilon+m_ec^2)+{\kappa^2\over r^2}\right]^{1/2}.
    \end{equation}

In our calculations we use the number node $n$=0 and $n$=1, for the
orbitals $1s_{1/2}$ and $2s_{1/2}$, respectively, $\kappa$ being -1.
Numerically, the eigenvalues of the discrete spectrum are obtained
by matching two numerical solutions of the Dirac equation: the
inverse solution that starts from the asymptotic conditions and the
direct one that starts at $r$=0.

The radial density of the bound state electron
w.f. on the nuclear surface is:

    \begin{equation}
        D^2_{n,\kappa}={1\over (m_ec^2)^3}\left({\hbar c\over a_0}\right)^3
        \left({a_0\over R_A}\right)^2
        \left[g_{n,\kappa}^2(R_A)+f_{n,\kappa}^2(R_A)\right], \label{media}
    \end{equation}
where $R_A=1.2A_0^{1/3}$ is given in fm, $a_0$ being the Bohr
radius. For the 1$s_{1/2}$ and 2$s_{1/2}$ electron orbitals, we use
$g_K^2=D^2_{0,-1}$ and $g_{L_1}^2=D^2_{1,-1}$, respectively.

For the $EC$ processes, the potential used to obtain the electron
w.f. reads

        \begin{equation}
            rV_{EC}(Z,r)=rV(Z,r)\phi(r),
        \end{equation}

\noindent and the charge number $Z=Z_0$ corresponds to the parent
nucleus. $V(Z,r)$ is negative.

The numerical solutions of the Dirac equation were obtained within
the power series method of Ref. \cite{Buh65}, by using similar
numerical algorithm as that of Refs. \cite{sal1,sal2}. The method is
able to provide numerical solutions of the Dirac equation for
central fields. We provide a grid with values of the potential for
different radial distances. The radial w.f. is expanded  in an
infinite power series that depends on the radial increment and the
potential values. The w.f. is calculated step by step in the mesh
points. The increment and the number of terms in the series
expansion determine the accuracy of the solutions. In our
calculations, the increment interval is 10$^{-4}$ fm and at lest 100
terms are taken into account in the series expansion. These values
exceed the convergence criteria of Ref. \cite{sal1}. To renormalize
the numerical solutions, we made use of the fact that at very large
distances, the behavior of the w.f. must approach that of the
Coulomb function. Therefore, the the amplitudes and the  the phase
shifts can be extracted by comparing the numerical solution and the
analytical ones. For discrete states, the asymptotic behavior of the
w.f. gives a guess for the inverse solutions. The eigenvalue is
obtained when the direct solutions and the inverse ones match each
other. We constructed an adequate procedure to find the bound states
of the electron up to an accuracy of 0.3 keV, or lower, by searching
solutions up to 130 keV binding energies. In this range of energies,
all the possible bound state energies are found. We calculated the
solutions starting outward from $r=0$ and inward from a very large
value of the radius $r$. The bound states should be obtained when
both solutions are equal in an intermediate point, for the two
components of the wave function. We found these energies by
interpolation. We selected the radial wave functions $f_{n,\kappa}$
and $g_{n,\kappa}$ that have same number of nodes $n=$0 or 1.

For the PSF computation, all integrals in Eq.~(\ref{dirac}) were
performed accurately with Gauss-Legendre quadrature in 32 points. We
calculated up to 49 values of the radial functions in the Q value
energy interval, that were interpolated with spline functions.

We also calculated the PSF for EC process using Eq.~(\ref{eckl}) but
employing essentially the formalism adopted by Ref. \cite{Mar70}.
Here we used the electron radial density (and density ratios) as
given in Table~2 of  \cite{Mar70}. Exchange corrections were taken
as unity. Binding energies were also taken from the same reference.

 \begin{figure*}
 \begin{center}
 \resizebox{0.90\textwidth}{!}{
   \includegraphics{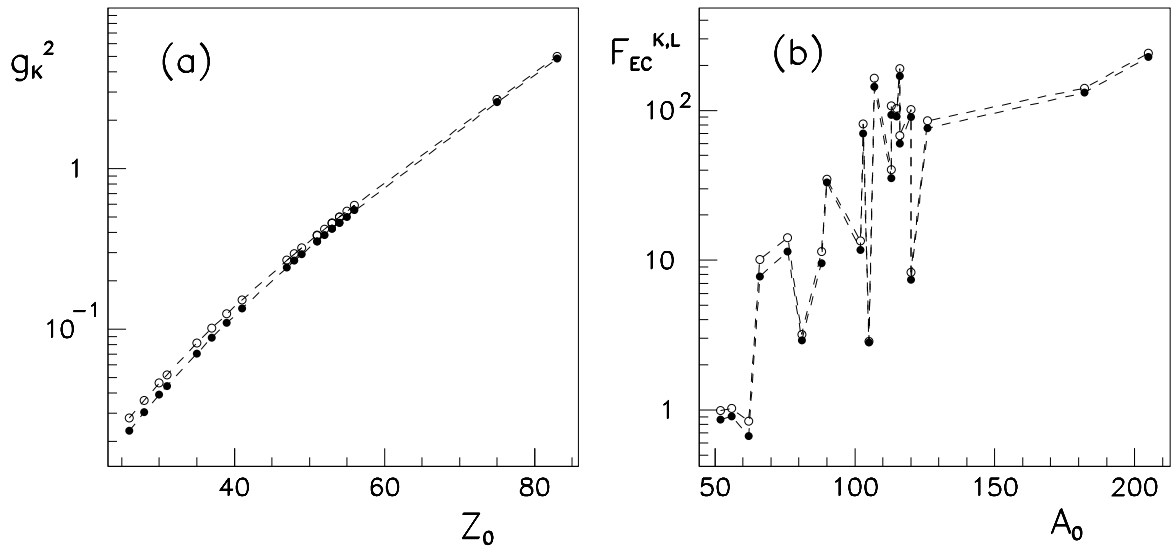}}
 \caption{(a) Electron density on the nuclear surface of the 1s$_{1/2}$
 state as a function of the atomic number of the parent nucleus. The
 values calculated with screening are displayed with filled circles
 and those without screening are plotted with empty symbols. The
 dashed lines are given to guide the eye. (b) Calculated PSF for
 EC as a function of the mass number of the parent
 nucleus. The filled circles display the calculation with screening,
 while the empty ones the calculation without screening.
 \label{psec}}
 \end{center}
 \end{figure*}

\section{\label{sec:level3}RESULTS AND DISCUSSION}

We perform PSF computations for the $\beta^+$ decay and the EC
process with the method described in the previous section, that we
call TW (This Work), for a large number of nuclei of experimental
interest.

For the $\beta^+$ decays we found previous PSF results  computed
with approximate methods \cite{TH73,Wil74}, for sixteen nuclei of
astrophysical interest. In Table I we display the PSF values for
these nuclei calculated with our new method (TW) and, for
comparison, the values taken from \cite{TH73,Wil74}. Also, we
present the PSF values computed by us using the recipe described in
Ref. \cite{Gov71}. All calculations were done with the $W_{0}$ value
indicated in \cite{Wil74}. One can see that the agreement between TW
results and the other results are in general under 1\%, except the
last two (heaviey) nuclei where the differences reach $\sim$ 3\%.

In Table II we display our PSF computed with the new method for few
heavy  nuclei, but for which we did not find previous results. For
comparison we computed the same PSF values with the recipe adopted
from  Ref. \cite{Gov71}. $W_{0}$-values were taken from \cite{Aud95}
for both sets of calculations. We found a rather good agreement
between the two sets of result, with differences within, generally,
a few percent. There was one exception, $^{105}$Ag, where the
difference was large ($\sim$ a factor 10). This is a case where the
$W_{0}$-value is very small (0.325 MeV), and might make our numerical
routine to be inaccurate at such small values. However, this discrepancy may not be so
significant, as long as the calculated PSF value is small enough to have
little contribution to the corresponding beta decay rates

In Table~III we preset our results for EC for the same set of
nuclei. The Q-values for positron decay were taken from Ref.
\cite{Wil74} for nuclei marked with $\star$. For the rest of nuclei
the Q-values were taken from Ref. \cite{Aud95}. Together with the
PSF values for EC,  the electron densities, $g_{K,L_1}$, their
ratios and the binding energies $\epsilon$ for the orbitals
$1s_{1/2}$ and $2s_{1/2}$ are also given in Table~III. We compare
the results performed with the new method (TW) with those calculated
using the recipe of Ref. \cite{Mar70}. For these transitions the
differences between the two sets of results are significantly larger
than for the positron decays, ranging from a few percent to about a
mammoth 35\%. We attribute these differences in the calculated PSF
values mainly due to electron densities, $g_K$, whose values,
calculated with the "old"' and "new"' methods, differ significantly
from each other. We also checked the influence of the screening
effect on the PSF values. We found that while for the positron
decays this effect is very small, for the EC transitions there is
some differences between the "`screened"' and "`un-screened" PSF
values. Fig. 1 shows this effect on the electron density, $g_K$ and
on the final PSF values. For small values of $Z$ the results without
screening give PSF values that are 10-15\% larger than those listed
in Table III. For heavier nuclei, these differences are only up to
2-3\%. The screening effect in PSF calculation is more important for
light nuclei and lead to a decrease in the PSF values up to 15\%.
Finally, in Table IV we present PSF values for EC transitions,
re-computed with up-dated Q-values taken from Ref. \cite{Aud12}. We
propose to use these new computed values of PSF for calculation of
$\beta$ decay rates.

\section{\label{sec:level4}SUMMARY AND CONCLUSION}

In summary, we constructed a new code for computing PSF values for
positron decays and EC processes. In our approach we get positron
free and electron bound w.f. by solving a Dirac equation with a
Coulomb-type potential, obtained from a realistic distribution of
protons in the daughter nuclei. The FNS and screening effects are
addressed as well by our new recipe. Using the same Q-values, we
compare our results with previous calculations where
electron/positron w.f. were obtained in an approximate way. For
positron decays the agreement with older results is quite good,
while for EC processes the differences between "new" and "old" PSF
values is as big as 35\%. We further found that the screening effect
is important for EC processes, specially for light nuclei, having an
impact up to 10-15\% on the calculated PSF values. Finally, using
our new method, we re-computed the PSF for all nuclei using up-dated
Q-values. We hope, these computed PSF values will prove useful in
more accurate estimations of the beta decay rates. We are currently
working on the impact of newly computed PSF values on $\beta$-decay
half-lives and hope to report our findings in near future.
\vspace{0.5in}

\textbf{Acknowledgments} S. Stoica, M. Mirea would like to acknowledge the support of the ANCSI-UEFISCDI through the project PCE-IDEI-2011-3-0318, Contract no. 58.11.2011.

J.-U. Nabi would like to acknowledge the support of the Higher
Education Commission Pakistan through the HEC Project No. 20-3099.

       %TABLES
       \onecolumngrid
       \clearpage
       \centering
       \setlength{\extrarowheight}{.5em}
       \begin{table}
        \centering \scriptsize \caption{Calculated phase space of
            $\beta^{+}$-decay (BP) compared with previous calculations. The
            value of maximum $\beta$-decay energy is taken from \cite{Wil74} for
            pure Fermi transitions. The last two columns show our calculated
            results.}\label{lightbp}

        \begin{tabular}{|c|c|c|c|c|c|}

            \hline
            Nucleus & $W_{0}$\cite{Wil74} & $F_{BP}$ \cite{TH73} & $F_{BP}$\cite{Wil74} & $F_{BP}$[TW] & $F_{BP}$\cite{Gov71}   \\
            & (MeV) & & & &  \\
            \hline
            $^{10}$C&0.8884                    &2.361          &2.361        &2.325       &2.326     \\
            $^{14}$O&1.8098                    &43.398         &43.378       &42.822      &42.814    \\
            $^{18}$Ne&2.383                    &136.83         &136.83       &135.19      &135.08    \\
            $^{22}$Mg&3.109                    &427.02         &426.88       &422.19      &421.51    \\
            $^{26}$Al&3.211                    &483.84         &483.68       &478.3       &477.43    \\
            $^{26}$Si&3.817                    &1036.8         &1035.9       &1025.51     &1023.059  \\
            $^{30}$S&4.439                     &1990.2         &1987.8       &1969.24     &1963.9    \\
            $^{34}$Cl&4.468                    &2014.7         &2013.4       &1993.13     &1987.4    \\
            $^{34}$Ar&5.021                    &3388.3         &3383.8       &3351.58     &3339.85   \\
            $^{38}$K&5.028                     &3346.9         &3344.9       &3312.82     &3300.54   \\
            $^{38}$Ca&5.620                    &5515.9         &5510.3       &5457.95     &5449      \\
            $^{42}$Sc&5.409                    &4533.5         &4531.7       &4490.19     &4462.21   \\
            $^{42}$Ti&5.964                    &7025.4         &7024.1       &6934.9      &6853.74   \\
            $^{46}$V&6.032                     &7285.9         &7284.2       &7186.04     &7091.9    \\
            $^{50}$Mn&6.609                    &10818          &10810        &10492.76    &10262     \\
            $^{54}$Co&7.227                    &15956          &15951        &14988.470   &14412.5   \\

            \hline

        \end{tabular}

       \end{table}

       %\clearpage
       \begin{table}
        \centering \scriptsize \caption{Calculated phase space of
            $\beta^{+}$-decay (BP) for heavy nuclei compared with the ones we
            calculated using recipe of \cite{Gov71}.}\label{heavybp}

        \begin{tabular}{|c|c|c|c|}
            \hline
            Nucleus & $W_{0}$ \cite{Aud95} & $F_{BP}[TW]$  & $F_{BP}$ \cite{Gov71} \\
            &   (Mev) &               &                         \\
            \hline
            $^{52}$Fe&1.3525             &8.3403           &8.4132       \\
            $^{56}$Ni&1.1109             &3.4439           &3.5250       \\
            $^{62}$Zn&0.5974             &0.2344           &0.2438       \\
            $^{66}$Ga&4.153              &1125.6442        &1132.5483    \\
            $^{76}$Br&3.9409             &835.1982         &843.3343     \\
            $^{81}$Rb&1.2161             &4.3222           &6.8878       \\
            $^{88}$Y&2.6006              &120.2644         &121.8624     \\
            $^{90}$Nb&5.0893             &2503.0555        &2533.7049    \\
            $^{102}$Cd&1.565             &11.2214          &11.5267      \\
            $^{103}$In&5.0005            &2100.3727        &2136.0153    \\
            $^{105}$Ag&0.325             &0.0102           &0.1127       \\
            $^{107}$Sb&6.837             &8528.5047        &8931.8197    \\
            $^{113}$Sb&2.8891            &168.1487         &172.0209     \\
            $^{113}$Te&5.048             &2124.1816        &2165.2927    \\
            $^{115}$I&4.7029             &1517.2376        &1549.2409    \\
            $^{116}$I&6.7547             &7913.1790        &8272.0244    \\
            $^{116}$Xe&3.235             &352.3565         &361.4082     \\
            $^{120}$Ba&3.98              &678.0918         &705.0294     \\
            $^{120}$Xe&0.5587            &0.1047           &0.1108       \\
            $^{126}$Cs&3.7731            &542.4653         &563.8184     \\
            $^{182}$Re&1.778             &16.123           &17.206       \\
            $^{205}$Bi&1.6835            &12.3984          &13.4576      \\
            \hline
        \end{tabular}

       \end{table}

       \clearpage

       \begin{table}
        \caption{Calculated phase space factors $F_{EC}$ for electron
            capture (assuming exchange corrections to be equal to 1). The value
            of maximum $\beta$-decay energy is taken from \cite{Wil74} for pure
            Fermi transitions. The electron densities, their ratios, and the
            binding energies $\epsilon$ are also provided for the orbitals
            1s$_{1/2}$ and 2s$_{1/2}$, including those given in \cite{Mar70}.
            Binding energies are given in units of $keV$.
        } \label{ectable}

        \begin{tabular}{|c|c|c|c|c|c|c|c|c|c|c|c|}
            \hline
            Nucleus &$Q_{\beta^+}$& $g_K^2$ &  $g_K^2$[TW] &$g_{L_1}^2/g_K^2$  & $g_{L_1}^2/g_K^2$[TW] &$\epsilon_K$ & $\epsilon_K$ [TW]&  $\epsilon_{L_1}$ & $\epsilon_{L_1}$[TW] &
            $F_{EC}^{K,L_1}$[TW] &
            $F_{EC}^{K,L_1}$       \\
            & (MeV)       &\cite{Mar70}& &      \cite{Mar70}       &             &   \cite{Mar70}             &                  &    \cite{Mar70}  &        &          & \cite{Mar70}                                    \\
            \hline
            $^{10}$C$^{\star}$&  1.9104       & 0.00031   &  0.00031  & 0.04930    & 0.02867 & 0.18790   & 0.62660 & 0.12600 & 0.01176 & 0.00703  &  0.00640     \\
            $^{14}$O$^{\star}$&  2.83186      & 0.00075   &  0.00065  & 0.05640    & 0.04420 & 0.40160   & 1.03733 & 0.02440 & 0.03251 & 0.03297  &  0.03786     \\
            $^{18}$Ne$^{\star}$& 3.405        & 0.00151   &  0.00118  & 0.05840    & 0.05794 & 0.68540   & 1.48302 & 0.03400 & 0.06659 & 0.08713  &  0.11005     \\
            $^{22}$Mg$^{\star}$& 4.131        & 0.00268   &  0.00199  & 0.06660    & 0.06811 & 1.07210   & 2.11143 & 0.06330 & 0.15721 & 0.218   &  0.29060     \\
            $^{26}$Al$^{\star}$& 4.2331       & 0.00344   &  0.00251  & 0.06990    & 0.07265 & 1.30500   & 2.40715 & 0.08940 & 0.14631 & 0.27558  &  0.39270     \\
            $^{26}$Si$^{\star}$& 4.839        & 0.00435   &  0.00312  & 0.07290    & 0.07661 & 1.55960   & 2.74689 & 0.11770 & 0.18077 & 0.47240  &  0.65060     \\
            $^{30}$S$^{\star}$&  5.461        & 0.00664   &  0.00467  & 0.07810    & 0.08342 & 2.14550   & 3.49498 & 0.18930 & 0.25934 & 0.90680  &  1.27140    \\
            $^{34}$Cl$^{\star}$& 5.4908       & 0.00807   &  0.00563  & 0.08040    & 0.08628 & 2.47200   & 3.91749 & 0.22920 & 0.30899 & 1.10727  &  1.56600     \\
            $^{34}$Ar$^{\star}$& 6.043        & 0.00970   &  0.00675  & 0.08240    & 0.08862 & 2.82240   & 4.33190 & 0.27020 & 0.36199 & 1.61130  &  2.28490     \\
            $^{38}$K$^{\star}$&  6.05         & 0.01156   &  0.00802  & 0.08440    & 0.09079 & 3.20600   & 4.77984 & 0.32630 & 0.41921 & 1.92311  &  2.73480    \\
            $^{38}$Ca$^{\star}$& 6.642        & 0.01367   &  0.00947  & 0.08620    & 0.09259 & 3.60740   & 5.25087 & 0.37710 & 0.48351 & 2.74237  &  3.90650     \\
            $^{42}$Sc$^{\star}$& 6.4311       & 0.01600   &  0.01113  & 0.08790    & 0.09430 & 4.03810   & 5.73657 & 0.43780 & 0.54865 & 3.02434  &  4.28930     \\
            $^{42}$Ti$^{\star}$& 6.986        & 0.01870   &  0.01300  & 0.08960    & 0.09579 & 4.49280   & 6.25222 & 0.50040 & 0.62068 & 4.17496  &  5.92320     \\
            $^{46}$V$^{\star}$&  7.0543       & 0.02170   &  0.01512  & 0.09100    & 0.09699 & 4.96640   & 6.78377 & 0.56370 & 0.69826 & 4.95575  &  7.02120     \\
            $^{50}$Mn$^{\star}$& 7.6311       & 0.02870   &  0.02016  & 0.09380    & 0.09920 & 5.98920   & 7.92722 & 0.69460 & 0.86703 & 7.74617  &  10.9103     \\
            $^{52}$Fe& 2.374        & 0.0328    & 0.0232    & 0.0950     & 0.0987  & 7.1120    & 8.5130  & 0.8461  &  0.958  & 0.859    & 1.2033     \\
            $^{54}$Co$^{\star}$& 8.2498       & 0.03730   &  0.02651  & 0.09620    & 0.10077 & 7.11200   & 9.14731 & 0.84610 & 1.05584 & 11.91799 &  16.6144     \\
            $^{56}$Ni& 2.136        & 0.0423    & 0.0303    &  0.0974    & 0.1013  & 8.3328    & 9.7882  & 1.0081  &  1.158  & 0.907    & 1.2580     \\
            $^{62}$Zn& 1.626        & 0.0538    & 0.0390    & 0.0995     & 0.1025  & 9.6586    & 11.157  &1.1936   &  1.380  & 0.675    & 0.9261     \\
            $^{66}$Ga& 5.175        & 0.0604    & 0.0410    & 0.1006     & 0.1029  & 10.3671   & 11.875  &1.2977   &  1.498  & 7.80     & 10.613     \\
            $^{76}$Br& 4.963        & 0.0935    & 0.0704    & 0.1035     & 0.1048  & 13.4737   & 15.000  &1.7820   &  2.021  & 11.45    & 15.162     \\
            $^{81}$Rb& 2.23815       & 0.1149    & 0.0883    & 0.1063     & 0.1080  & 15.1997   & 16.690  &2.0651   &  2.263  & 9.069    & 11.744     \\
            $^{88}$Y& 3.6226        & 0.1402    & 0.1091    & 0.1080     & 0.1174  & 17.0384   & 18.450  &2.3725   &  2.438  & 9.528    & 12.114     \\
            $^{90}$Nb& 6.111        & 0.170     & 0.1344    & 0.1098     & 0.1059  & 18.9856   & 20.421  & 2.6977  &  2.994  & 33.17    & 41.975     \\
            $^{102}$Cd& 2.587       & 0.319     & 0.2663    & 0.1159     & 0.1102  & 26.7112   & 28.044  &4.0180   &  4.351  & 11.66    & 14.019     \\
            $^{103}$In& 6.050       & 0.348     & 0.2930    & 0.1168     & 0.1116  & 27.9399   & 29.232  &4.2375   &  4.548  & 71.05    & 84.541     \\
            $^{105}$Ag& 1.345       & 0.293     & 0.2423    & 0.1150     & 0.1086  & 25.5140   & 26.864  &3.8058   &  4.161  & 2.816    & 3.4256     \\
            $^{107}$Sb& 7.920       & 0.413     & 0.3526    & 0.1187     & 0.1096  & 30.4912   & 31.726  &4.6983   &  5.095  & 146.5    & 172.43     \\
            $^{113}$Sb& 3.913       & 0.413     & 0.3516    & 0.1187     & 0.1096  & 30.4912   & 31.726  &4.6983   &  5.095  & 35.38    & 41.804     \\
            $^{113}$Te& 6.070       & 0.449     & 0.3844    & 0.1196     & 0.1113  & 31.8138   & 33.041  &4.9392   &  5.314  & 93.70    & 109.93     \\
            $^{115}$I&  5.729       & 0.488     & 0.4121    & 0.1205     & 0.1124  & 33.1694   & 34.345  &5.1881   &  5.542  & 91.54    & 106.40     \\
            $^{116}$I&  7.780       & 0.488     & 0.4215    & 0.1205     & 0.1124  & 33.1694   & 34.345  &5.1881   &  5.542  & 169.3    & 196.75     \\
            $^{116}$Xe& 4.450       & 0.529     & 0.4609    & 0.1215     & 0.1123  & 34.5644   & 35.705  &5.4528   &  5.822  & 60.15    & 69.410     \\
            $^{120}$Ba& 5.00        & 0.623     & 0.5496    & 0.1234     & 0.1130  & 37.4406   & 38.514  &5.9888   &  6.375  & 90.65    & 103.51     \\
            $^{120}$Xe& 1.617       & 0.529     & 0.4599    & 0.1215     & 0.1123  & 34.5644   & 35.705  &5.4528   &  5.821  & 7.72     & 8.9482     \\
            $^{126}$Cs& 4.824       & 0.574     & 0.501     & 0.1224     &  0.112  & 35.9846   & 37.111  &5.7143   &   6.128 & 76.88    & 88.697     \\
            $^{182}$Re& 2.800       & 2.69      & 2.593     & 0.1448     &  0.128  & 71.6764   & 72.491  &12.5267  &   13.26 & 22.86    & 24.152     \\
            $^{205}$Bi& 2.708       & 4.88      & 4.837     & 0.1561     &  0.138  & 90.5259   & 91.373  &16.2370  &   17.25 & 228.17   & 233.83     \\

            \hline
        \end{tabular}
       \end{table}

       \clearpage

       \begin{table}
        \caption{Calculated phase space factors $F_{EC}$ for electron
            capture, with  Q-values from \cite{Aud12}.
        } \label{ecnew}
        \begin{tabular}{|c|c|c|c|c|c|c|}
            \hline
            Nucleus & $Q_{EC}$\cite{Aud12}  & $F_{EC}$[TW] &  $F_{EC}$\cite{Mar70} & $F_{BP}$[TW]& $F_{BP}$\cite{Gov71}\\
            & (MeV)  & &&& \\
            \hline
            $^{10}$C&    3.64613 &  0.07318  & 2.33265 & 226.780     & 226.834 \\
            $^{14}$O&    5.14131 &  0.21794  & 0.12483 & 1644.76     & 1643.41 \\
            $^{18}$Ne&   4.44215 &  0.27831  & 0.18733 & 677.970     & 677.912 \\
            $^{22}$Mg&   4.77904 &  0.61616  & 0.39020 & 995.887     & 995.685 \\
            $^{26}$Al&   4.00231 &  0.62642  & 0.35240 & 343.398     & 343.658 \\
            $^{26}$Si&   5.06645 &  0.51788  & 0.71694 & 1339.344    & 1339.30 \\
            $^{30}$S&    6.13834 &  1.14585  & 1.61931 & 3805.276    & 3803.16 \\
            $^{34}$Cl&   5.48869 &  1.10642  & 1.57889 & 1994.797    & 1995.09 \\
            $^{34}$Ar&   6.05858 &  1.61963  & 2.31915 & 3410.133    & 3409.96 \\
            $^{38}$K&    5.91093 &  1.83565  & 2.64042 & 2917.839    & 2918.62 \\
            $^{38}$Ca&   6.73867 &  2.82284  & 4.07367 & 5924.355    & 5929.26 \\
            $^{42}$Sc&   6.42269 &  3.01643  & 4.33609 & 4470.946    & 4471.87 \\
            $^{42}$Ti&   7.01275 &  4.20702  & 6.05196 & 7100.190    & 7130.06 \\
            $^{46}$V&    7.04865 &  4.94781  & 7.11022 & 7175.692    & 7209.06 \\
            $^{50}$Mn&   7.63042 &  7.74479  & 11.0705 & 10516.941   & 10744.5 \\
            $^{52}$Fe&   2.37330 &  0.8584  & 1.22082 & 14942.286   & 15765.2 \\
            $^{54}$Co&   8.24017 &  11.89015 & 16.8306 & 8.354       & 8.43206 \\
            $^{56}$Ni&   2.13175 &  0.9029  & 1.27259 & 3.444       & 3.49486 \\
            $^{62}$Zn&   1.61859 &  0.6687  & 0.93259 & 0.234       & 0.24131 \\
            $^{66}$Ga&   5.17225 &  7.7902  & 10.7797 & 1125.644    & 1131.60 \\
            $^{76}$Br&   4.96024 &  11.439  & 15.4388 & 835.295     & 841.531 \\
            $^{81}$Rb&   2.23696 &  2.9044  & 3.84415 & 4.321       & 4.41092 \\
            $^{88}$Y&    3.62067 &  9.5180  & 12.3759 & 120.264     & 121.864 \\
            $^{90}$Nb&   6.10809 &  33.141  & 42.9337 & 2502.372    & 2526.00 \\
            $^{102}$Cd&  2.58562 &  11.652  & 14.4027 & 11.221      & 11.5468 \\
            $^{103}$In&  6.01928 &  70.333  & 15.7203 & 2099.402    & 2133.61 \\
            $^{105}$Ag&  1.34679 &  2.8233  & 3.53114 & 0.01027     & 1.12362 \\
            $^{107}$Sb&  7.85483 &  144.059 & 174.745 & 8528.505    & 8918.59 \\
            $^{113}$Sb&  3.90909 &  35.311  & 42.9919 & 168.122     & 172.036 \\
            $^{113}$Te&  6.06682 &  93.601  & 113.234 & 2124.182    & 2162.53 \\
            $^{115}$I&   5.72192 &  91.3148 & 109.531 & 1509.977    & 1547.75 \\
            $^{116}$I&   7.77260 &  168.959 & 202.635 & 7930.046    & 8250.78 \\
            $^{116}$Xe&  4.44315 &  59.963  & 71.4659 & 354.467     & 361.241 \\
            $^{120}$Ba&  4.99761 &  90.562  & 106.993 & 685.518     & 703.098 \\
            $^{120}$Xe&  1.57992 &  7.3638  & 8.82085 & 0.105       & 0.11187 \\
            $^{126}$Cs&  4.79256 &  75.871  & 90.4835 & 542.400     & 555.411 \\
            $^{182}$Re&  2.79851 &  131.273 & 145.184 & 16.123      & 17.2282 \\
            $^{205}$Bi&  2.70412 &  227.499 & 247.263 & 12.415      & 13.4798 \\
            \hline
        \end{tabular}
       \end{table}

       \clearpage

    \end{document}